\documentclass[a4paper,12pt]{revtex4}

\usepackage{graphicx}
\usepackage{dcolumn}
\usepackage{bm}
\usepackage{newtxtext,newtxmath}
\usepackage[T1]{fontenc}
\usepackage{ae,aecompl}
\usepackage[mathscr]{euscript}

\let\mathscr\relax 
\usepackage[scr]{rsfso}
\usepackage{enumitem}
\usepackage{tabularx, booktabs, makecell}
\usepackage{siunitx}
\usepackage{tikz}
\usetikzlibrary{shapes,arrows}
\newcommand{\civ}{C\,\textsc{iv}}
\newcommand{\mgii}{Mg\,\textsc{ii}}
\newcommand{\hi}{H\,\textsc{i}}

\begin{document}

\title{Lines of sight through randomly oriented flattened spheroids; quasar absorption cloud sizes}

\author{John K. Webb\footnote{See postscript regarding collaborators.}}
\affiliation{Clare Hall, University of Cambridge, Herschel Rd, Cambridge CB3 9AL, UK.}

\begin{abstract}
\noindent{\bf Abstract} \\ Randomly oriented flattened spheroids have been used to describe a broad range of astrophysical phenomena. Here we use this geometric approach to derive equations representing lines of sight through quasar absorption clouds to constrain cloud sizes.
\end{abstract}

\maketitle

\section{Introduction}

The first direct association between Lyman-$\alpha$ forest quasar absorption lines and galaxies near the Earth-quasar sightline was reported in \cite{Lanzetta1995}. Those observations revealed the enormity of galactic gaseous envelopes at intermediate redshifts; most galaxies at $z \lesssim 1$ are surrounded by {\hi} halos $\approx 160 h^{-1}$ kpc radius and these halos explain a significant fraction of the Lyman-$\alpha$ forest at intermediate redshift. 

The discoveries motivated similar explorations into the extent of heavier element halos \citep{Carswell1995} with similarly surprising results; using 14 galaxy/absorber pairs, the covering factor of {\civ} gas was found to extend to radii of $\sim 100 h^{-1}$ kpc \citep{Chen2001}. Subsequent studies confirmed these discoveries and revealed the detailed physics and environments of young galaxies over a broader redshift range. Deep imaging studies in the fields of very high redshift quasars pushed faint-object identification to sensitive limits, finding that {\civ} absorbers at $z \sim 5-6$ may arise in Lyman-$\alpha$ emitters or possibly in low mass galaxies \cite{Diaz2014}.

The properties of {\mgii} absorption in extended galactic halos \citep{Tinker2008} was investigated using double quasar sightlines \citep{Rogerson2012} and \cite{Chen2010} used a large sample of quasar-galaxy pairs to study the galactic absorption signature imprinted on the spectrum of the background quasar, deriving detailed mass and kinematic information from {\mgii} transitions. In a recent detailed study, \cite{Lundgren2021} analyse 54 {\mgii} absorption systems towards nine quasars at $z \sim 2$ and find that 89\% of the systems in the range $0.64<z<1.6$ can be associated with a galaxy, at impact parameter less than $<200$ kpc.

Large-scale surveys show that galaxies frequently reside in cosmic web filaments and that their formation mechanisms generate spin alignments with the filaments \citep{Tempel2013, Dubois2014}.

\section{A geometric parameterisation} \label{sec:geometry}

With even a small amount of thermal Doppler broadening, the profiles of absorbing atoms in terms of line density are dominated by the broadening rather than the rotation \citep{Pitts1997}. This identifies the velocity at which there is a maximum density of absorbing atoms. In this paper we take this velocity to be the one which will be observed in the centre of the absorption line. There may be some error in this because we have not considered the line formation in detail but this should not affect the general character of our conclusions. In particular it must introduce less inaccuracy than that involved in our assumption that the absorbing atoms are smoothly distributed in space with only thermal Doppler broadening.

\begin{figure}
\begin{centering}
\begin{tikzpicture}
\draw[->] (0, 0) -- (2.5, 0) 
node[right]{\footnotesize $y$};
\draw[->] (0, 0) -- (0, 2.5)
node[above]{\footnotesize $z$};
\draw[->,blue] (0, 0) -- (2.3, -1.35)
node[above]{\footnotesize $x'$};
\draw[->,blue] (0, 0) -- (1.1, 0.5)
node[above]{\footnotesize $y'$};
\draw[->,blue] (0, 0) -- (1.2, 2.3)
node[above]{\footnotesize $z'$};
\draw[->] (0, 0) -- (1.8, -1.8)
node[above]{\footnotesize $x$};
\draw[-,red] (0, 0) -- (-0.6, 0.7)
node[left,red]{\footnotesize $P$};
\node[below,red] at (-0.3, 0.3) {\footnotesize $r$}; 
\node[below] at (0.2, 1) {\footnotesize $\theta$};
\draw[->] (0:0.7) arc (0:-45:0.7);
\node[below] at (0.8, -0.05) {\footnotesize $\phi$};
\filldraw[red] (-0.6, 0.7) circle(1.5pt);
\draw[->] (-0.6, 3.5) node[above]{\footnotesize {\sc Quasar}} -- (-0.6, -3.5)
node[below]{\footnotesize {\sc Observer}};
    \begin{scope}[rotate=-30, xscale=2.2, yscale=0.55, opacity=0.4]
      \coordinate (O) at (0,0);
      \shade[ball color=gray!10!] (0,0) circle (1) ;
    \end{scope}
\end{tikzpicture}
\caption{Absorbing cloud represented by randomly oriented oblate spheroid, showing the two coordinate systems. The semi-major axis, $a$ (not illustrated), lies along the $x'$ axis. Section \ref{sec:geometry} gives details.}
\label{fig:geometry}
\end{centering}
\end{figure}
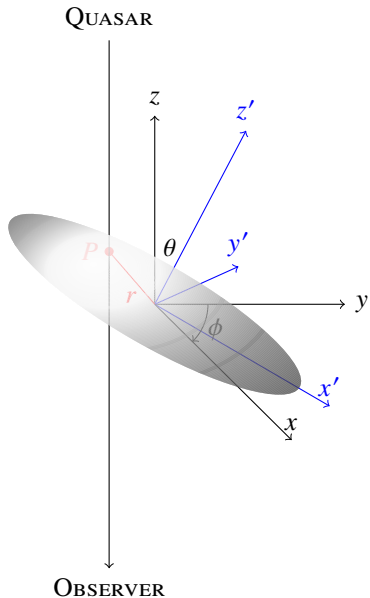

Assuming the absorbing clouds to be halos of disc galaxies, we approximate them as flattened rotating oblate spheroids whose volume density is constant on similar concentric spheroids. The rotation velocity rises rapidly near the centre of the disc to give an approximately flat rotation velocity at large radii. The surface density of the disc is a truncated exponential which vanishes at the edge of a spheroid of semi major axis $a$; the volume density vanishes at the surface of that spheroid. The coordinate systems of the problem are shown in Figure \ref{fig:geometry}. The orientation of the spheroid and the position of the line of sight are defined by three quantities $\theta, \phi$ and $r$. $\theta$ is the angle between the minor axis of the spheroid and the line of sight, $\phi$ completes the definition of the orientation of the spheroid and $r$ is the perpendicular displacement of the line of sight from the centre of the spheroid. All possible orientations and displacements can then be studied by giving random values between 0 and 1 for the three quantities $\cos \theta, 2\phi/\pi$ and $r^2/a^2$.\\

\section{Single sightlines through absorbing gas clouds}

\cite{Chen2001} reported observations of twelve CIV clouds with measured impact parameters relative to the centres of galaxies with which they are believed to be associated. Eight of the clouds have impact parameters less than 75 kpc, two between 75 kpc and 150 kpc and two between 150 kpc and 225 kpc. What is perhaps more important than these numbers is the fractional number of intersections at a given impact parameter. We now discuss what can be learnt about the probable sizes of the clouds. Two points can be made immediately. The observed clouds must be at least as large as the observed impact parameters. However, the number of large clouds must be overestimated in the observations because they present a large surface area to the radiation from quasars. We discus two possibilities. The first is that the clouds are spherical. The second is that they are highly flattened.

\subsection{Spherical clouds}

Suppose that there is a distribution of spherical clouds such that $n(r)dr$ have radii between $r$ and $r+dr$. Then the number of lines of sight which intersect the clouds out to impact parameter $r$ is proportional to the total cross-sectional area out to that radius, which is 
\begin{equation}
A(r)=\int_{0}^{r}n(r^{\prime })\pi r^{\prime}{}^{2}dr^{\prime}+\int_{r}^{\infty}n(r^{\prime })\pi r^{2}dr^{\prime}.
\label{eq:Ar}
\end{equation}

The fractional number of sight lines at radius $r$ which intersect a cloud is 
\begin{equation}
f(r)=\frac{\int_{r}^{\infty }n(r^{\prime })dr^{\prime}}{\int_{0}^{\infty}n(r^{\prime })dr^{\prime}}.
\label{eq:fr1}
\end{equation}
A crude approximation to observational data may be made using Equation \ref{eq:fr1}, from which we can give expressions for $n(r)$,
\begin{equation}
f(r)=[1+r/r_{0})^{2}]^{-1}
\label{eq:fr2}
\end{equation}
and 
\begin{equation}
f(r)=\exp (-0.7r/r_{0}),
\label{eq:fr3}
\end{equation}
where in each case $r_{0}=37.5$ kpc. The corresponding expressions for $n(r)$ are 
\begin{equation}
n(r)\propto (r/r_{0})[1+(r/r_{0})^{2}]^{2}
\label{eq:a5}
\end{equation}
and 
\begin{equation}
n(r)\propto \exp (-0.7r/r_{0}).
\label{eq:a6}
\end{equation}

Obviously these expressions for $n(r)$ are quite different but there is much less variation in the implied average radii of the clouds. For the two distributions they are 59 kpc and 54 kpc respectively. The data will obviously accept variations on the functional forms of equations \eqref{eq:fr2} and \eqref{eq:fr3}. Thus tolerable fits are obtained with $r/r_0$ in Equation \eqref{eq:fr2} replaced by $0.9r/r_0$ or $1.1r/r_0$ and with $0.7r/r_0$ in Equation \eqref{eq:fr3} replaced by $0.65r/r_0$ or $0.75r/r_0$. The resulting range in the average cloud radius is 50 kpc to 65 kpc.

In fact the fits which we have made are not quite correct. The data are binned in ranges of 75 kpc of impact parameter although they are plotted at the mid-points of the bins. We will not repeat our analysis but it is possible, for example, to take a function like Equation \eqref{eq:fr2} for $f(r)$ with an arbitrary coefficient before $r/r_0$ and to ask for what value of this coefficient the average of $f(r)$ over the bins best fits the data. We find that we require the spheres to be between ten and twenty per cent larger than in the previous fit. The upshot of this discussion in that the average radii of spherical clouds required to fit the data probably lie in the range 50 kpc to 75 kpc. Each of the distributions of equations \eqref{eq:a5} and \eqref{eq:a6} has a small number of much larger clouds, which show up in the observations.

One final comment is that no precise meaning can be applied to the size of a cloud. What is being discussed here is the size of the clouds down to the lowest column density of CIV observed in \cite{Chen2001}.

\subsection{Flattened clouds}

Suppose an infinitely flattened disc has major axis $a$. It can be shown that the number of possible intersections for radii between $r$ and $r + dr$ for $r \leq a$, averaged over all orientations of the disc is $p(r,a)da$, where 
\begin{equation}
p(r,a) = 2\pi r \left[1 - \frac{2}{\pi} \sin^{-1} \frac{r}{a} \right].
\label{eq:pra}
\end{equation}
This is normalised so that $p/2\pi r = 1$ if all intersections are possible. Suppose that there is a distribution of major axes of discs $n(a)da$ such that $\int_0^{\infty} n(a)da = 1$. Define $N(a) = \int_o^a n(a^{\prime})da^{\prime}$, so that $N(\infty) = 1$. Then, if $P(r)dr$ is the total number of possible intersections, 
\begin{align}
P(r) & = \int_r^{\infty} p(r,a)n(a)da  \nonumber \\
& = 2\pi r \int_r^{\infty} \left(1 - \frac{2}{\pi} \sin^{-1} \frac{r}{a} \right) (dN/da) da  \nonumber \\
& = 2 \pi r \left[N(\infty) - N(r) - \frac{2}{\pi} \int_r^{\infty} \sin^{-1} \frac{r}{a} (dN/da)da \right]  \nonumber \\
& = 2 \pi r \left[N(\infty) - N(r) - \left[ \frac{2}{\pi} N \sin^{-1}  \frac{r}{a} \right]_r^{\infty}\right] + \nonumber \\
& \qquad\qquad\qquad\qquad\qquad 2 \pi r \left[ \frac{2}{\pi} \int_r^{\infty} Nd \left(\sin^{-1} \frac{r}{a} \right) \right]  \nonumber \\
& = 2 \pi r \left[ 1 - \frac{2r}{\pi} \int_r^{\infty} \frac {N(a)da}{a(a^2 - r^2)^{1/2}} \right].
\label{eq:Pr}
\end{align}
So far this is quite general. Now suppose that the discs have the same distribution of $a$ that we have assumed for the spheres. In the first case take 
\begin{align}
& n(a) = 2a/a_0^2[1 + (a/a_0)^2]^2  \nonumber \\
& N(a) = 1 - 1/[1 + (a/a_0)^2].
\label{eq:Na}
\end{align}
Then 
\begin{align}
& P(r) = 2 \pi r \left[ 1 - \frac{2r}{\pi} \int_r^{\infty} \frac{da}{a(a^2 - r^2)^{1/2}} \nonumber \right] + \nonumber \\
& \qquad\qquad\qquad\qquad 2 \pi r \left[ \frac{2r}{\pi} \int_r^{\infty} \frac{da}{a(a^2 - r^2)[1 + (a/a_0)^2]} \right].
\label{eq:Pr2}
\end{align}
The first integral in the brackets is readily evaluated and it has the value 1. Two successive changes of variable enable the second integral to be evaluated to give the result \begin{equation}
P(r)/2\pi r = 1 - (r/a_0)/[1 + (r^2/a_0^2)]^{1/2} \equiv f(r).
\label{eq:Pr3}
\end{equation}
The expression \eqref{eq:Pr3} can now be compared with the observational data and the best value of $a_0$ for a fit obtained. Using the data of \cite{Chen2001}, a good fit is obtained with $r_0 = 0.63a_0$. This suggests that, if the clouds are flattened discs, they need to be on average about 60 per cent larger than the spheres ($a_0 = 1.59r_0$). The argument can be repeated for 
\begin{align}
& n(a) = \exp(-a/a_0)/a_0  \nonumber \\
& N(a) = 1 - \exp(-a/a_0).
\label{eq:Nr2}
\end{align}
In that case 
\begin{align}
P(r) & = 4r^2 \int_r^{\infty} \frac{\exp(-a/a_0)da}{a(a^2 - r^2)^{1/2}} \nonumber \\ 
& \equiv 4rF(r).
\label{eq:Pr4}
\end{align}
The substitution $a = rt$ gives 
\begin{equation}
F(r) = \int_1^{\infty} \frac{\exp(-rt/a_0)dt}{t(t^2 - 1)^{1/2}}
\label{eq:Fr4}
\end{equation}
and a well-known expression for a zero order modified Bessel function yields 
\begin{equation}
dF/dr = -(1/a_0)K_0(r/a_0).
\label{eq:dFdr}
\end{equation}
From Equation \eqref{eq:Fr4} it is easy to see that $F(0) = \pi/2$ and this then gives 
\begin{equation}
P(r)/2\pi r = 1 - \frac{2}{\pi} \int_0^r K_0(r/a_0)dr/a_0 \equiv f(r).
\label{eq:Pro2pi}
\end{equation}
The integral in Equation \eqref{eq:Pro2pi} has to be evaluated numerically and the resulting $f(r)$ can be fitted to the observations of \cite{Chen2001} to find the best value of $a_0$. This is given approximately by $a_0 = r_0/0.44$. Again this gives discs about sixty percent larger than the spheres. 

\subsection{Inferences}

We conclude from the limited data available that the average radii of the CIV clouds down to the lowest column density at which they are observed lies between 50 and 75 kpc if they are spherical and between 80 and 120 kpc if they are highly flattened discs. There must be a significant spread of size. This is obvious from the observations if they are spherical but is equally true for flattened discs.

\section{Double sightlines through absorbing clouds}

There are a number of absorption line clouds which are detected in the spectra of two nearby quasars on the sky. For each sight line column densities and redshifts can be observed. The difference in redshifts provides a line of sight velocity displacement for the two point of the cloud being observed. The observed displacement of the two lines of sight on the sky together with the cloud redshift gives a value for the physical displacement of the two lines of sight which depends on the value of Hubble's constant and on the other parameters of the cosmological model, specifically the density parameter.

To study the properties of the distribution of two lines of sight through a single cloud, we must first consider random values of $\cos \theta$ and for each of these select a pair of random values of $2\phi/\pi$ and $r^2/a^2$. The symmetry of the expression for column density means that we need only consider values of each of $\theta$ and $\phi$ between 0 and $\pi/2$. We used the same values when we considered velocity profiles but that meant that we could not determine whether the velocity along a particular sight line runs red shifted or blue shifted relative to the disc centre. This did not matter because we were only concerned with the velocity profiles and the magnitude of the displacement from the central velocity. Here things are different. Each calculation represents four pairs of values of $\phi$, two corresponding to velocities in the same direction and two in the opposite direction. Finally each of the four cases gives a different line of sight displacement on the sky. Thus each set of random numbers provides four cases.

It is clear that the maximum value of $\Delta r/a$ is 2 and that the maximum value of $\Delta v_z/v_o$ is less than 2, where $v_o$ is the asymptotic value of the rotation velocity at infinity. Our first interest is to determine the distribution of these quantities for random orientations and a variety of density profiles and velocity profiles. For the cases studied, a majority of values of $\Delta v_z/v_o$ are below 0.5 and a majority of the values of $\Delta r/a$ are less than 1.0. If we further demand that the column density along each sight line exceeds some threshold value, so that there is a strong chance that both sight lines would in fact have been identified, rather obviously the likely value of $\Delta r/a$ is reduced.

As both $a$ and $v_o$ are free parameters in our models, we cannot make an immediate comparison with what is observed. However the present observations are not incompatible with the properties of rotating discs with acceptable values of $a$ and $v_o$. Our calculations suggest that the number of observations should be large otherwise very misleading conclusions could be drawn from a small random sample of cases. Our models would be tested much more seriously if any cloud with two quasar sight lines were to be identified with a galaxy because then the displacements of the two lines of sight from the centre of the galaxy would be separately observable and at least some reasonable approximation of the orientation of the galaxy might be deduced. 

Consider first the four cases which correspond to each pair of values of our random variables. For a given value of $\cos \theta$ the shape of the spheroid projected perpendicular to the line of sight is an ellipse with axes $a$ and $(a^2 \cos 2\theta + c^2 \sin 2\theta)^{1/2}$ where $c$ is the minor axis of the spheroid and we shall use the notation $p \equiv a^2/c^2$. Points in this ellipse are defined by values of $r$/a and $\phi$. Suppose we have two pairs of values $(r_1/a, \phi_a)$ and $(r_2/a, \phi_2)$. Choose $0 \leq \phi_1, \phi_2 \leq \pi/2$. Then there are four values of $\phi$ which can be associated with $\phi_1$ to give the the column densities and $|\Delta v_z|/v_o$. These are $\phi_2, \pi-\phi_2, \pi+\phi_2$ and $2\pi-\phi_2$. The first two correspond to the same sign of $\Delta v_z$ as that of $\phi_2$ and hence the difference in the two velocities must be chosen. The other two have the opposite sign of $\Delta v_z$ so that the sum of the two velocities must be chosen. All four cases give different values of $|r_1 - r_2|$ which can be calculated from the cosine rule for the four triangles.

We identified two possibilities in one of which the line of sight velocity is monotonic across of the line of sight through the cloud and in the other of which there is a turning value in the velocity. In this latter case there is an infinite number density of HI atoms per unit velocity at the turning value in the absence of any broadening mechanism. Even a relatively small amount of broadening produces a more more or less symmetrical profile but it scarcely changes the velocity at the peak of the line from that of the turning value. In such cases we have simply taken the turning value velocity to be the appropriate value of $\Delta v_z$. When there is no turning value the broadening mechanism can change the position of the peak of the profile and to make certain that we determine it accurately we have had to take a fine grid of values of $v_z$ to make certain that we obtain the correct value of $\Delta v_z$.

If the calculation is performed with the slight modification just described we obtain two random values of the column density for each random pair $(\theta, \phi_1, r_1), (\theta, \phi_2, r_2)$. With the two column densities are associated two values of $\Delta v_z$ (the sums and differences of the individual $\Delta v_z$) and, with each value of $\Delta v_z$, two values of $|r_1 - r_2|$.

The rotation curve studied is
\begin{equation}
v = v_o r' /(r' + \epsilon a)
\label{eq:v}
\end{equation}
where $r \equiv (x'^2 + y'^2)^{1/2}$ is the distance from the minor axis of the spheroid. Calculations have been performed for $\epsilon = 0.05$ and $\epsilon = 0.1$. These values correspond to a rotation velocity at the edge of the disc of $0.952 v_o$ and $0.909 v_o$ respectively. The surface number density of the disc has the form
\begin{equation}
N = \frac{N_o [e^{-kr'/a} + (kr'/a) e^{-3kr'/a} - e^{-k}]}{(1 - e^{-k})}
\label{eq:N}
\end{equation}
and we have made calculations for $k = 6.4$ and $k = 20$. For $k = 6.4$ there are few effective e-foldings of the column density between the centre and edge for the disc which means that in most cases, if one column density is large enough to be observed, the other will be clear. That is not the case for $k = 20$, so that we have considered the effect on the statistics of our results by demanding that both calculated column densities exceed some prescribed profile. Most of our calculations have $p = 100$ which corresponds to an axis ratio of 10, but we have done one calculation with $p = 1000$ and one with $p = 10$ to observe the effect of changing the axis ratio. We have  also used $p = 1$, which is not appropriate for a flattened rotating cloud, simply to see what extreme changes in properties can be obtained.

\section*{Postscript}

This incomplete paper grew from collaborative work that commenced when Roger Tayler, Eric Pitts, and John Webb were working together in the Astronomy Centre at the University of Sussex. The work was near completion when Roger Tayler's death halted the project, which was then shelved for many years. Original handwritten notes by RJT were transcribed, modified, and included in the text provided here. Additional text places that original work in the context of a more developed observational field. Although the work was never finished, rather than not seeing the light of day, it is hoped that the theoretical models described in this paper (non-peer reviewed and arXiv-only) may be useful either the quasar absorption field, or in some other context.

\bibliographystyle{mnras}
\bibliography{references}

\newcommand{\noop}[1]{}
\begin{thebibliography}{}
\makeatletter
\relax
\def\mn@urlcharsother{\let\do\@makeother \do\$\do\&\do\#\do\^\do\_\do\%\do\~}
\def\mn@doi{\begingroup\mn@urlcharsother \@ifnextchar [ {\mn@doi@}
  {\mn@doi@[]}}
\def\mn@doi@[#1]#2{\def\@tempa{#1}\ifx\@tempa\@empty \href
  {http://dx.doi.org/#2} {doi:#2}\else \href {http://dx.doi.org/#2} {#1}\fi
  \endgroup}
\def\mn@eprint#1#2{\mn@eprint@#1:#2::\@nil}
\def\mn@eprint@arXiv#1{\href {http://arxiv.org/abs/#1} {{\tt arXiv:#1}}}
\def\mn@eprint@dblp#1{\href {http://dblp.uni-trier.de/rec/bibtex/#1.xml}
  {dblp:#1}}
\def\mn@eprint@#1:#2:#3:#4\@nil{\def\@tempa {#1}\def\@tempb {#2}\def\@tempc
  {#3}\ifx \@tempc \@empty \let \@tempc \@tempb \let \@tempb \@tempa \fi \ifx
  \@tempb \@empty \def\@tempb {arXiv}\fi \@ifundefined
  {mn@eprint@\@tempb}{\@tempb:\@tempc}{\expandafter \expandafter \csname
  mn@eprint@\@tempb\endcsname \expandafter{\@tempc}}}

\bibitem[\protect\citeauthoryear{{Carswell}}{{Carswell}}{1995}]{Carswell1995}
{Carswell} R.~F.,  1995, \mn@doi [Nature] {10.1038/374500a0}, 374, 500–501

\bibitem[\protect\citeauthoryear{{Chen}, {Lanzetta}  \& {Webb}}{{Chen}
  et~al.}{2001}]{Chen2001}
{Chen} H.-W.,  {Lanzetta} K.~M.,   {Webb} J.~K.,  2001, \mn@doi [ApJ]
  {10.1086/321537}, \href
  {https://ui.adsabs.harvard.edu/abs/2001ApJ...556..158C} {556, 158}

\bibitem[\protect\citeauthoryear{{Chen}, {Helsby}, {Gauthier}, {Shectman},
  {Thompson}  \& {Tinker}}{{Chen} et~al.}{2010}]{Chen2010}
{Chen} H.-W.,  {Helsby} J.~E.,  {Gauthier} J.-R.,  {Shectman} S.~A.,
  {Thompson} I.~B.,   {Tinker} J.~L.,  2010, \mn@doi [ApJ]
  {10.1088/0004-637X/714/2/1521}, \href
  {https://ui.adsabs.harvard.edu/abs/2010ApJ...714.1521C} {714, 1521}

\bibitem[\protect\citeauthoryear{{D{\'\i}az}, {Koyama}, {Ryan-Weber}, {Cooke},
  {Ouchi}, {Shimasaku}  \& {Nakata}}{{D{\'\i}az} et~al.}{2014}]{Diaz2014}
{D{\'\i}az} C.~G.,  {Koyama} Y.,  {Ryan-Weber} E.~V.,  {Cooke} J.,  {Ouchi} M.,
   {Shimasaku} K.,   {Nakata} F.,  2014, \mn@doi [MNRAS]
  {10.1093/mnras/stu914}, \href
  {https://ui.adsabs.harvard.edu/abs/2014MNRAS.442..946D} {442, 946}

\bibitem[\protect\citeauthoryear{{Dubois} et~al.,}{{Dubois}
  et~al.}{2014}]{Dubois2014}
{Dubois} Y.,  et~al., 2014, \mn@doi [MNRAS] {10.1093/mnras/stu1227}, \href
  {https://ui.adsabs.harvard.edu/abs/2014MNRAS.444.1453D} {444, 1453}

\bibitem[\protect\citeauthoryear{{Lanzetta}, {Bowen}, {Tytler}  \&
  {Webb}}{{Lanzetta} et~al.}{1995}]{Lanzetta1995}
{Lanzetta} K.~M.,  {Bowen} D.~V.,  {Tytler} D.,   {Webb} J.~K.,  1995, \mn@doi
  [ApJ] {10.1086/175459}, \href
  {https://ui.adsabs.harvard.edu/abs/1995ApJ...442..538L} {442, 538}

\bibitem[\protect\citeauthoryear{{Lundgren} et~al.,}{{Lundgren}
  et~al.}{2021}]{Lundgren2021}
{Lundgren} B.~F.,  et~al., 2021, \mn@doi [ApJ] {10.3847/1538-4357/abef6a},
  \href {https://ui.adsabs.harvard.edu/abs/2021ApJ...913...50L} {913, 50}

\bibitem[\protect\citeauthoryear{{Pitts} \& {Tayler}}{{Pitts} \&
  {Tayler}}{1997}]{Pitts1997}
{Pitts} E.,  {Tayler} R.~J.,  1997, \mn@doi [MNRAS] {10.1093/mnras/288.2.457},
  \href {https://ui.adsabs.harvard.edu/abs/1997MNRAS.288..457P} {288, 457}

\bibitem[\protect\citeauthoryear{Rogerson \& Hall}{Rogerson \&
  Hall}{2012}]{Rogerson2012}
Rogerson J.~A.,  Hall P.~B.,  2012, \mn@doi [MNRAS]
  {10.1111/j.1365-2966.2011.20317.x}, 421, 971

\bibitem[\protect\citeauthoryear{{Tempel}, {Stoica}  \& {Saar}}{{Tempel}
  et~al.}{2013}]{Tempel2013}
{Tempel} E.,  {Stoica} R.~S.,   {Saar} E.,  2013, \mn@doi [MNRAS]
  {10.1093/mnras/sts162}, \href
  {https://ui.adsabs.harvard.edu/abs/2013MNRAS.428.1827T} {428, 1827}

\bibitem[\protect\citeauthoryear{{Tinker} \& {Chen}}{{Tinker} \&
  {Chen}}{2008}]{Tinker2008}
{Tinker} J.~L.,  {Chen} H.-W.,  2008, \mn@doi [ApJ] {10.1086/587432}, \href
  {https://ui.adsabs.harvard.edu/abs/2008ApJ...679.1218T} {679, 1218}

\makeatother
\end{thebibliography}

\end{document}